\documentclass[10pt]{article}
\usepackage{graphicx}
\usepackage{latexsym,amssymb,amsmath,amscd,amsfonts,mathabx}
\newtheorem{theorem}{Theorem}[section]

\newtheorem{lemma}[theorem]{Lemma}
\newtheorem{cor}[theorem]{Corollary}
\newtheorem{remark}[theorem]{Remark}

\newcommand{\vf}{\varphi}

\newcommand{\bbP}{\mathbb P}
\newcommand{\eproof}{\rule{0.2cm}{0.2cm}}

\newcommand{\Cdot}{\raisebox{-0.25ex}{\scalebox{1.5}{$\cdot$}}}

\newcommand{\ds }{ \displaystyle }
\newcommand{\re}{\mathbb{R}}

\begin{document}

\title{\bf Fractional Fokker-Planck-Kolmogorov equations associated with stochastic differential equations in a bounded domain}

\author{Sabir Umarov}
\date{}
\maketitle

\begin{center}
{\it University of New Haven, Department of Mathematics, \\300 Boston Post Road, West Haven, CT 06516, USA}
\end{center}
%\ead{sumarov@newhaven.edu}
\vspace{10pt}
%\begin{indented}
%\item[]September 2014
%\end{indented}

\begin{abstract}
This paper is devoted to the fractional generalization of the Fokker-Planck equation associated with a stochastic differential equation in a bounded domain. The driving process of the stochastic differential equation is a L\'evy process subordinated to the inverse of L\'evy's mixed stable subordinators.  The Fokker-Planck equation is given through the general Waldenfels operator, while the boundary condition is given through the general Wentcel's boundary condition. As a fractional operator a distributed order differential operator with a Borel mixing measure is considered. In the paper fractional generalizations of the Fokker-Planck equation are derived and the existence of a unique solution of the corresponding initial-boundary value problems is proved.

\end{abstract}

% Uncomment for PACS numbers
%\pacs{05.40-a, 02.50-r, 05.10Gg}
%
% Uncomment for keywords
\vspace{2pc}

\section{Introduction}\label{sec:1}

The Fokker-Planck equation, including fractional versions, play an important role in the modeling of various random processes. Its applications can be found in all branches of statistical physics \cite{Bogolyubov,Gardiner,Kampen}, in quantum mechanics \cite{Bogolyubov,Shizgal}, in biology \cite{Bressloff,Presse}, in finance \cite{Melnikov,Kampen}, just to mention a few. The most of these applications assume an idealistic model in which a random quantity may take values in the whole $d$-dimensional space $\re^d.$ Fractional generalizations of the Fokker-Planck equation in this case are obtained in works \cite{Chechkin,Gajda,HKU-1,HU,MetzlerKlafter00,Weron}; see also references therein. The connection of a wide class of fractional Fokker-Planck equations with their associated stochastic differential equation driven by time-changed L\'evy processes was first studied in paper \cite{HKU-1}. Chapter 7 of the book  \cite{Umarov_book} presents all the details of such a connection between fractional order Fokker-Planck-Kolmogorov (FPK for short) equations and the stochastic differential equations in the whole space $\mathbb{R}^d.$ 

What concerns stochastic processes in a bounded domain, they are also spread out broadly, an example of which is a diffusion in a bounded region. The essential difference of the stochastic process in a bounded domain from the case of stochastic processes in $\re^d$ is the influence of the boundary or near boundary processes to the whole picture.  There is a rich literature on the stochastic processes and the associated (non fractional) FPK equations in a bounded domain; see recent monograph by Taira \cite{Taira}. 

In the present paper we will discuss fractional generalizations of FPK equations associated with the stochastic differential equations in a bounded domain with a smooth boundary. The fractional diffusion and associated Fokker-Planck equation in a bounded domain was studied in papers \cite{Agrawal,Burch,MMM-1,MMM-2,Tomovski} in various particular cases. Our goal in this paper is to study general initial-boundary value problems describing stochastic processes undergoing inside a bounded region, as well as on its boundary. We also discuss initial-boundary value problems for fractional FPK equations associated with stochastic differential equations driven by fractional Brownian motion in a bounded domain. Note that fractional FPK equation in the whole space $\mathbb{R}^d$ associated with stochastic processes driven by fractional Brownian motion was considered in papers \cite{BC,Biagini,HU,HKU-2}.

For the reader's convenience we start the discussion with a well known particular cases of the influence of the boundary and related boundary conditions. Consider a stochastic differential equation
\begin{equation}
\label{sde_111}
dX_t = b(t,X_t) dt + \sigma(t,X_t) dB_t, \quad X_0=x,
\end{equation}
where $x \in \mathbb{R}^d$ is a random vector independent of $m$-dimensional Brownian motion $B_t$, 
and the mappings 
$$b: (0,\infty) \times \mathbb{R}^d \to \mathbb{R}^d, \quad \sigma : (0,\infty) \times \mathbb{R}^d \to \mathbb{R}^{d \times m}$$ 
satisfy Lipschitz and linear growth conditions. Namely, there exist positive constants $C_1$ and $C_2$ such that for all $x, y \in \mathbb{R}^d$  the inequalities
\begin{equation}
\label{lip_1}
\sum_{k=1}^d |b_k(t,x)-b_k(t, y)| +\sum_{k=1}^d \sum_{\ell=1}^m |\sigma_{k,\ell}(t,x)-\sigma_{k,\ell}(t,y)| \le C_1|x-y|,
\end{equation}
and
\begin{equation}
\label{lg_1}
\sum_{k=1}^{d} |b_k(t, x)| +\sum_{k=1}^d \sum_{\ell=1}^m |\sigma_{k,\ell}(t, x)| \le C_2 (1+|x|)
\end{equation}
hold.
The stochastic differential equation \eqref{sde_111} is understood as
\begin{equation}
\label{sde_112}
X_t=X_0+\int\limits_0^t b(s, X_s)ds+\int\limits_0^t \sigma(s, X_s)dB_s,
\end{equation}
with the second integral in the sense of It\^o.

The backward Kolmogorov equation associated with SDE \eqref{sde_111}  
 in terms of the density function $u(t,x)$ of $X_t^x=(X_t|X_0=x)$ is
\begin{equation}
\label{sde_115}
\frac{\partial u(t,x)}{\partial t} = \mathcal{A}(t) u(t,x), \quad t>0, x \in \re^d,
\end{equation}
where $t$-dependent differential operator $A(t)$ is defined in the form
\begin{equation}
\label{sde_120}
\mathcal{A}(t) = \sum_{k=1}^d b_k(t, x) \frac{\partial}{\partial x_k} +  \sum_{j,k=1}^d a_{j,k}(t, y)\frac{\partial^2}{\partial x_j \partial x_k},
\end{equation} 
with $(d \times d)$-matrix-function $\{a_{j,k}(t,x), \ j,k=1\dots d\},$ coinciding with  $\frac{1}{2} \sigma(t,x) \sigma^{T}(t,x).$ Here $\sigma^{T}(t,x)$ is the matrix transposed to the matrix-function $\sigma(t,x).$
The corresponding Fokker-Planck equation, or forward Kolmogorov equation, has the form
\begin{equation}
\label{sde_211}
\frac{\partial u(t,x)}{\partial t} = \mathcal{A}^{\ast}(t) u(t,x), \quad t>0, x \in \re^d,
\end{equation}
where $\mathcal{A}^{\ast}$ is the formally adjoint operator to $\mathcal{A}(t).$ Both equations \eqref{sde_115} and \eqref{sde_211} are accompanied with the initial condition
\begin{equation}
\label{sde_116}
u(0,x)=u_0(x), \quad x\in \re^d,
\end{equation}
where $u_0(x)$ is the density function of the initial vector $X_0.$
Unifying the terminology used in these two equations we call the pair of equations \eqref{sde_115} and \eqref{sde_211} {\it the FPK equation}, the term used throughout the current paper.

If the solution process $X_t$ of the stochastic differential equation in \eqref{sde_111} 
is allowed to change only in a bounded region $\Omega \subset \re^d$ with a smooth boundary $\partial \Omega,$ then the probability $\bbP(X_t \in \re^d \setminus \Omega)$ of $X_t$ being out of the region $\Omega$ is zero. The associated FPK equation in this case needs to be supplemented by boundary conditions. In order to  see in what form the boundary conditions are given, we need to introduce the notion of
{\it probability current,} a $d$-dimensional vector field $\ds {\bf \Phi(t,x)},$ components of which are defined as % 
\[
\Phi_k(t,x)=  b_k(t,x) p(t,x) - \sum_{j=1}^d \frac{\partial}{\partial y_j} \Big[ a_{j,k}(t,x) p(t,x) \Big], \quad k=1,\dots,d.
\]
Using the probability current, one can write forward FPK equation \eqref{sde_211} in the form of a conservation law:
\[
\frac{\partial p(t,x)}{\partial t} + \sum_{k=1}^d \frac{\partial \Phi_k(t,x)}{\partial x_k}=0,
\]
or, the same,
\[
\frac{\partial p(t,x)}{\partial t} + \nabla \ \Cdot \ {\bf \Phi (t,x)}=0,
\] 
where $\nabla =(\frac{\partial}{\partial x_1}, \dots, \frac{\partial}{\partial x_d})$ is the gradient operator and  the symbol ``$\Cdot$" means the dot product of two vector objects.

Let $S$ be a $(d-1)$-dimensional hyper-surface in $\Omega,$ and ${\bf n}_x, \ x\in S,$ be an outward normal to $S$ at the point $x \in S.$ Then the total flow of probabilities through the hyper-surface $S$ can be calculated by the surface integral
\[
\int_S {\bf \Phi} (t,x) \cdot {\bf n}_x dS.
\]
Two boundary conditions are common in the study of stochastic processes in a bounded region: 
\begin{enumerate}
\item[(a)] {\it reflecting} boundary condition, and 
\item[(b)] {\it absorbing} boundary condition.
\end{enumerate} 
The reflecting boundary condition means that there is no probability flow across the boundary, leading to the condition
\begin{equation}
\label{refl_500}
{\bf \Phi}(t,x) \cdot {\bf n}_x =0, \quad x \in \partial \Omega.
\end{equation}
In this case $X_t$ will stay in the region $\Omega$ forever and be reflected when $X_t$ reaches the boundary $\partial \Omega.$ The absorbing boundary means that the probability for $X_t$ being on the boundary is zero, that is $\bbP(X_t \in \partial \Omega)=0,$ leading to the condition
\begin{equation}
\label{absorb_500}
p(t,x)=0. \quad x \in \partial \Omega.
\end{equation}
In this case $X_t$ will be absorbed by boundary as soon as $X_t$ reaches the boundary.

Let a boundary operator $\mathcal{B}$ be defined in the form
\begin{equation}
\label{bound_op_000}
\mathcal{B}\vf(x)=\mu(x) \vf(x) + \nu(x) \frac{\partial \vf}{\partial {\bf n}_x}, \quad  x\in \partial \Omega,
\end{equation}
where the functions $\mu(x)$ and $\nu(x)$ are continuous on the boundary $\partial \Omega,$ satisfying some physical conditions discussed below, and ${\bf n}_x$ is the outward normal at point $x \in \partial \Omega.$
In order to define operators that will be used in the FPK equations associated with the SDEs in a bounded domain we introduce the following spaces:
\[
C^2_{\mathcal{B}}(\Omega):= \{\vf \in C^2(\Omega): \mathcal{B} \vf(x) =0, x \in \partial \Omega \},
\]
with the boundary operator $\mathcal{B}$ defined in \eqref{bound_op_000}.
Introduce the operator $A_{\mathcal{B}}(t)$ on the space $C^2_{\mathcal{B}}(\Omega)$: 
\begin{align*}
\label{FPK_bound_500}
A_{\mathcal{B}}(t):= \Big\{ A(t) \vf(x)  &= \sum_{k=1}^d b_k(t, x) \frac{\partial \vf(x)}{\partial x_k} +\sum_{j,k=1}^d a_{j,k}(t, x)\frac{\partial^2 \vf(x)}{\partial x_j \partial x_k}, \quad x \in \Omega \Big\}.
%\\ 
%\mathcal{B} \vf (x)_{|x \in \partial \Omega} &=0 \Big\}.
\end{align*} 
In other words the operator $A_{\mathcal{B}}(t)$ formally is the same as the operator $\mathcal{A}(t)$ defined in \eqref{sde_120}, but with the domain $\ds Dom(A_{\mathcal{B}}(t))=C^2_{\mathcal{B}}(\Omega)$ for each fixed $t >0.$
The operator $\ds A_{\mathcal{B}}$ is linear and maps the space $C^2_{\mathcal{B}}(\Omega)$ to $C(\Omega).$ %(see this and other properties of the operator $A_{\mathcal{B}}$ in Chapter 4).

Both boundary conditions for stochastic processes in a domain with absorbing or reflexive boundaries discussed above can be reduced to the boundary condition 
$$ 
\ds \mathcal{B} p(t,x) = 0, \quad t>0, \ x \in \partial \Omega.
$$ 
Therefore,
the forward and backward FPK equations in the case of bounded domain in terms of the density function $u(t,x)$ of $X_t^x=(X_t|X_0=x)$ can be formulated with the help of the operator $A_{\mathcal{B}}.$ Namely, the backward FPK equation %
is given by
\[
\frac{\partial u(t,x)}{\partial t} = A_{\mathcal{B}} u(t,x), \quad t>0, x \in \Omega,
\]
and the forward FPK equation is given by
\[
\frac{\partial u(t,x)}{\partial t} = {A}_{\mathcal{B}}^{\ast}(t) u(t,x), \quad t>0, x \in \Omega,
\]
where $A_{\mathcal{B}}^{\ast}$ is the formally adjoint operator to ${A}_{\mathcal{B}}.$ % 

The paper is organized as follows.   Section 2  provides necessary  preliminary information on general (integer order) FPK equations through the Waldenfels operator and Wentcel's boundary conditions. The corresponding stochastic processes represent solutions of associated stochastic differential equations in a bounded domain with combined continuous diffusion, jump, and other relevant phenomena. The main results of the paper is in Section 3. In this section we derive initial-boundary value problems for fractional order FPK equations. The associated stochastic processes represent solutions of stochastic differential equations in a bounded domain with an appropriate time-changed driving processes. In this section we also prove the existence of a unique solution to the obtained initial-boundary value problems.

%%%%%%%%%%%%%%%%%%%%%
%%%%%%%%%%%%%%%%%%%%%

\section{Preliminaries and auxiliaries} 
\label{Preliminaries}

The FPK equation associated with a SDE with drift and diffusion coefficients independent on the time variable $t$ in the explicit form is given by the initial-boundary value problem 
\begin{align}
\frac{\partial u(t,x)}{\partial t} &=  \sum_{j=1}^d b_j(x) \frac{\partial u(t,x)}{\partial x_j} + \sum_{i,j=1}^d a_{i,j}(x) \frac{\partial^2 u(t,x)}{\partial x_i \partial x_j}  
\quad t >0, \ x \in \Omega,    
\label{bdom_500}
 \\
\mathcal{B} u(t,x^{\prime}) & \equiv  \mu(x^{\prime}) \frac{\partial u (t, x^{\prime})}{\partial {\bf n}} +\nu(x^{\prime})u(t,x^{\prime}) =0, \quad t>0, \ x^{\prime} \in \partial \Omega, 
\label{bdom_501}
\\
u(0,x) & = \delta_0(x), \quad x \in \Omega,
\label{bdom_502}
\end{align}
where $b_j(x), j=1,\dots,d,$ are drift coefficients and $a_{i,j}(x), i,j=1,\dots,d,$ are diffusion coefficients; the functions $\mu(x^{\prime})$ and $\nu(x^{\prime})$ are continuous functions defined on the boundary $\partial \Omega.$  It is well known that the stochastic process solving SDE corresponding to the FPK equation in \eqref{bdom_500}-\eqref{bdom_502} is continuous. Such a stochastic process represents a diffusion process of a Markovian particle. 
 
FPK equation \eqref{bdom_500}-\eqref{bdom_502} does not take into account jumping events and some specific phenomena (diffusion on the boundary, jumps on the boundary or into the domain, viscosity, etc.) which may occur on the boundary. The general case with jump and viscosity effects involve pseudo-differential operators in the FPK equation and in the boundary condition. The theory of pseudo-differential operators was first developed by Kohn and Nirenberg \cite{KohnNirenberg} and H\"ormander \cite{Hormander}. An operator $A$ is called pseudo-differential with a symbol $\sigma_A(x,\xi),$ if 
\begin{equation}
\label{psdo0}
A f(x) =\frac{1}{(2\pi)^d} \int_{\re^d} \sigma_A(x, \xi) e^{-ix \xi} \tilde{f}(\xi) d\xi,
\end{equation}
where $\tilde{f}(\xi)$ is the Fourier transform of $f:$ 
\[
\tilde{f}(\xi)=\int_{\re^n} f(x)e^{i x \xi} dx.
\]
The symbols of pseudo-differential operators in \cite{KohnNirenberg,Hormander} are smooth and satisfy some growth conditions as $|\xi| \to \infty.$ However, symbols of pseudo-differential operators considered in the present paper are not smooth. We refer the reader to works \cite{BN,Jacob,Umarov_book} where appropriate classes of symbols are studied.

For simplicity we assume that $\Omega \subset \re^n$ is a bounded domain with a smooth boundary $\partial \Omega \subset \re^{n-1}.$  In the general case the (backward) FPK equation can be formulated as follows:
\begin{align}
\frac{\partial u(t,x)}{\partial t}  = A(x,D) u(t,x), \quad & t>0, \ x \in \Omega,
\label{bound_500} 
\\
\mathcal{W}(x^{\prime}, D) u(t, x^{\prime})  =0, \quad & t>0, \ x^{\prime} \in \partial \Omega,
\label{bound_501}
\\
u(0,x) =u_0(x), \quad & x \in \Omega,
\label{bound_502}
\end{align} 
where $A(x,D)$ is a second order Waldenfels operator acting on the space of twice differentiable functions defined on $\Omega,$ that is
\begin{align}
\hspace{-.5cm}A (x,D) & \vf(x)  = c_0(x) \vf(x) +\sum_{i=1}^n b_i(x) \frac{\partial \vf(x)}{\partial x_i} + 
\sum_{i j=1}^n a_{ij}(x) \frac{\partial^2 \vf(x)}{\partial x_i \partial x_j} \notag
\\
&+
\int\limits_{\Omega}
\left[ \vf(y)  - \phi(x, y) \left( \vf(x) - \sum_{i=1}^n(y_i - x_i) \frac{\partial \vf(x)}{\partial y_i} \right) \right] \nu(x, dy), \quad x \in \Omega,
\label{bound_5}
\end{align}
and $\mathcal{W}(x^{\prime}, D)$ is a boundary pseudo-differential operator defined on the space of twice differentiable functions defined on $\partial \Omega$ through the local coordinates $x^{\prime}=(x_1,\dots,x_{n-1}) \in \partial \Omega$ (see \cite{Taira}):
\begin{align}
\mathcal{W}(x^{\prime},D) 
= Q(x^{\prime},D) + \mu(x^{\prime}) \frac{\partial }{\partial{\bf n}} - \delta(x^{\prime}) A(x^{\prime},D) + \Gamma_1 (x^{\prime},D) + \Gamma_2 (x^{\prime},D),
\label{bound_503}
\end{align}
with (pseudo)-differential operators
\begin{align}
Q&(x^{\prime},D)  \vf(x^{\prime})  =    \sum_{j,k=1}^{n-1} \alpha_{j,k}(x^{\prime}) \frac{\partial ^2\vf(x^{\prime})}{\partial x_j \partial x_k} + \sum_{j=1}^{n-1} \beta(x^{\prime}) \frac{\partial \vf(x^{\prime})}{\partial x_j}+\gamma(x^{\prime})\vf(x^{\prime}),
\label{bound_504}
\\
\Gamma_1 &(x^{\prime}, D)  \vf (x^{\prime}) \notag
\\
&=  \int_{\partial \Omega} \left[\vf(y^{\prime}) - \tau_1(x^{\prime}, y^{\prime}) \left( \vf(x^{\prime}) + \sum_{k=1}^{n-1} (y_k-x_k) \frac{\partial \vf(x^{\prime})}{\partial x_k}  \right)  \right] \nu_1(x^{\prime}, dy^{\prime}), 
\label{bound_505}
\end{align}
and
\begin{align}
\label{bound_506}
\Gamma_2 & (x^{\prime}, D)  \vf (x^{\prime}) \notag \\
&= \int_{\Omega} \left[\vf(y) - \tau_2(x^{\prime}, y) \left( \vf(x^{\prime}) + \sum_{k=1}^{n-1} (y_k-x_k) \frac{\partial \vf(x^{\prime})}{\partial x_k}  \right)  \right] \nu_2(x^{\prime}, dy).
\end{align}
Here $\tau_j(x^{\prime}, y^{\prime}), \ j=1,2,$ are local unit functions and $\nu_j(x^{\prime},dy^{\prime}), j=1,2,$ are L\'evy kernels satisfying some conditions indicated below, and $u_0$ is the density function of the initial state $X_0.$ If the initial state $X_0=0,$ then $u_0(x)=\delta_0(x).$ The boundary condition \eqref{bound_501} is called  a {\it second order Wentcel's boundary condition},  \index{Wentcel boundary condition} 
to credit Wentcel's contribution \cite{Wentcel} to the theory of diffusion processes.

One can work with coefficients $a_{ij}(x), i.j=1.\dots,n,$ $b_j(x), j=1,\dots, n,$ $c_0(x),$ and the symbol $\sigma_{A}(x,\xi)$ of the operator $A(x,D)$ in \eqref{bound_5} satisfying mild conditions similar to Lipschitz \eqref{lip_1} and linear growth condition \eqref{lg_1}. However, in this paper, for simplicity, we assume the following conditions:
\begin{enumerate}
\item[(i)] $a_{ij}(x) \in C^{\infty}(\Omega) \cap C(\overline{\Omega})$ and $a_{ij}(x) = a_{ji}(x)$ for all $i,j =1,\dots,n,$ and $x\in \overline{\Omega},$ and  there exists a constant $a_0>0$ such that
\[
\sum_{i, j=1}^n a_{ij}(x) \xi_i \xi_j \ge a_0 |\xi|^2, \quad x \in \overline{\Omega}, \ \xi \in \re^n;
 \]
 \item[(ii)] $b_j(x) \in C^{\infty}(\Omega) \cap C(\overline{\Omega}), \ j=1,\dots,n;$
 \item[(iii)] $c_0(x) \in C^{\infty}(\Omega) \cap C(\overline{\Omega}),$ and $c_0(x) \le 0$ in $\overline{\Omega};$
 \item[(iv)]  the local unity function $\phi(x,y)$ and the kernel function $\nu(x,dy)$ are such that the symbol $\sigma_A(x,\xi)$ of the operator $A(x,D)$ belongs to the class of symbols $S(\Omega \times \re^d)= C(\overline{\Omega} \times \re^d) \cap C^{\infty}(\Omega \times (\re^d \setminus G)),$ where $G \in \re^d$ is a set of $d$-dimensional measure  zero; see \cite{Umarov_book} for the theory of pseudo-differential operators with such a nonregular symbols.
\end{enumerate}
Equation \eqref{bound_500} describes a diffusion process accompanied by jumps in $\Omega$ with the drift vector $(b_1(x),\dots,b_n(x))$ and diffusion coefficient  defined by the matrix-function $a_{i, j}(x),$ and jumps governed by the L\'evy measure $\nu(x, \cdot).$ 
We assume additionally that the condition
\begin{enumerate}
\item[(C1)] 
\hspace{2cm}$ \ A(x,D) [1(x)] = c_0(x) + \int\limits_{\Omega} [1-\phi(x,y)] \nu(x, dy) \le 0, \quad x\in \Omega,$
\end{enumerate}
is fulfilled to ensure that  the jump phenomenon from $x \in \Omega$ to the the outside of a neighborhood of $x$ is ``dominated'' by the absorption phenomenon at $x$ (see \cite{Taira} for details). 

The coefficients $\alpha_{ij}(x^{\prime}), i.j=1.\dots,n,$ $\beta_j(x^{\prime}), j=1,\dots, n,$ and 
$\gamma(x^{\prime})$ 
of the operator $Q(x,D)$ in \eqref{bound_5} satisfy the following conditions:
\begin{enumerate}
\item[(a)] $\alpha_{ij}(x^{\prime}) \in C^{\infty}(\partial{\Omega})$ and $\alpha_{ij}(x^{\prime}) = \alpha_{ji}(x^{\prime})$ for all $i,j =1,\dots,n-1,$ and $x^{\prime} \in \partial {\Omega},$ and  there exists a constant $\alpha_0>0$ such that
\[
\sum_{i, j=1}^{n-1} \alpha_{ij}(x^{\prime}) \xi_i \xi_j \ge \alpha_0 |\xi|^2, \quad x^{\prime} \in \partial {\Omega}, \ \xi \in \re^{n-1};
 \]
 \item[(b)] $\beta_j(x^{\prime}) \in C^{\infty}(\partial \Omega), \ j=1,\dots,n;$
 \item[(c)] $\gamma(x^{\prime}) \in C^{\infty}(\partial \Omega),$ and $\gamma(x^{\prime}) \le 0$ in $\partial {\Omega}.$
 \end{enumerate}

 The symbols $\sigma_{\Gamma_1}(x^{\prime},\xi)$ and  $\sigma_{\Gamma_2}(x^{\prime},\xi)$ of the boundary pseudo-differential operators $\Gamma_1(x^{\prime}, D)$ and $\Gamma_2(x^{\prime}, D)$ in equations \eqref{bound_505} and \eqref{bound_506} satisfy the following condition: 
 
 \begin{enumerate}
 \item[(d)]  the local unity functions $\phi_k(x,y), \ k=1,2,$ and the kernel function $\nu_k(x,dy), \ k=1,2,$ are such that the symbols $\sigma_{\Gamma_k}(x^{\prime},\xi)$ of operators $\Gamma_k(x^{\prime},D), k=1,2,$ belong to the class of symbols $S(\partial \Omega, \re^{d})= %C(\partial {\Omega} \times \re^{n-1}) \cap 
 C^{\infty}(\partial \Omega \times (\re^{d} \setminus G_0)),$ where $G_0 \in \re^d$ is a set of $d$-dimensional measure zero. 
\end{enumerate}

The boundary condition \eqref{bound_501} with the operator $\mathcal{W}(x,D)$ defined in equations \eqref{bound_503}-\eqref{bound_506} describes a combination of continuous diffusion and jumping processes undergoing on the boundary, as well as jumps from the boundary into the region, and the viscosity phenomenon near the boundary. Namely, 
the term
\[
\sum_{j,k=1}^{n-1} \alpha_{j,k}(x^{\prime}) \frac{\partial^2 u(t, x^{\prime})}{\partial x_j \partial x_k} + \sum_{j=1}^{n-1} \left(\beta(x^{\prime}) \frac{\partial u(t,x^{\prime})}{\partial x_j}  +b(x^{\prime}) \frac{\partial u(t,x^{\prime})}{\partial x_j} \right)
\]
governs the diffusion process on the boundary, the term $\gamma(x^{\prime})  u(t, x^{\prime})$ is responsible for the absorption phenomenon at the point $x^{\prime} \in \partial \Omega,$ the term
$\mu(x^{\prime}) \frac{\partial }{\partial{\bf n}_{x^{\prime}}}$ expresses the reflexion phenomenon at $x^{\prime} \in \partial \Omega,$ the term $\delta(x^{\prime}) A(x^{\prime},D) u(t, x^{\prime})$ expresses the viscosity near $x^{\prime} \in \partial \Omega,$ and the terms $\Gamma_1 (x^{\prime}, D)  u (t, x^{\prime})$ and $\Gamma_2 (x^{\prime}, D)  u (t, x^{\prime})$ govern jump processes on the boundary and jump processes from the boundary into the region, respectively.  
We assume that the condition
\begin{align*}
\hspace{-3.8cm} \mbox{(C2)} \hspace{2cm} \mathcal{W}(x^{\prime},D) [1(x)] &=  \gamma(x^{\prime}) 
+ \int\limits_{\partial \Omega} [1-\tau_1(x^{\prime},y^{\prime})] \nu_1(x^{\prime}, dy^{\prime} )
\\
&+
\int\limits_{\Omega} [1-\tau_2(x^{\prime},y)] \nu_2(x^{\prime}, dy)
\le 0, \quad x^{\prime} \in \partial \Omega,
\end{align*}
is fulfilled to ensure that  the jump phenomenon from $x^{\prime} \in \partial \Omega$ to the outside of a neighborhood of $x^{\prime}$ on the boundary $\partial \Omega$ or inward to the region $\Omega$ is ``dominated'' by the absorption phenomenon at $x^{\prime}.$

We also assume the following transversality condition of the boundary operator $\mathcal{W}:$
\[
\hspace{-6.2cm} \mbox{(C3)}  \hspace{2cm} \int_{\Omega} t(x^{\prime}, y) dy = \infty, \quad \mbox{if} ~\mu(x^{\prime})= \delta(x^{\prime})=0.
\]

Fractional order FPK equations use fractional derivatives, and in more general case, fractional distributed order operators. Below we briefly recall the definitions and some properties of these operators referring the reader to the sources \cite{Kochubei,MSh,Umarov_book,UG} for details. By definition, the Caputo-Djrbashian derivative of order $0<\beta<1$ is
defined by
\begin{equation}
\label{Caputo}
  \,_{\tau} \mathcal{D}_{\ast}^{\beta} g(t) =
    {
      \frac{1}{\Gamma(1-\beta)}
      \int_{\tau}^{t} \frac{g^{\prime}(s) d s}{(t-s)^{\beta}}, \quad   \tau <t      
      },
\end{equation}
where $\tau \ge 0$ is an initial point and $\Gamma(z), \ z \in \mathbb{C} \setminus \{\dots, -2, -1, 0\},$ is  Euler's gamma function. We assume that the function $g(t), \ t \ge 0,$ is a differentiale function. We write
$\mathcal{D}_{\ast}^{\beta}$ when $\tau=0.$ Making use of the fractional integration
operator of order $\alpha >0$
\[
_{\tau} J^{\alpha} f(t)= \frac{1}{\Gamma(\alpha)} \int_{\tau}^t (t-s)^{\alpha-1}f(s)ds, 
\] 
we can write $_{\tau} \mathcal{D}_{\ast}^{\beta}$ also  in the operator form 
\[
_{\tau} \mathcal{D}_{\ast}^{\beta}= \ _{\tau}J^{1-\beta}\frac{d}{dt}.
\] 
In our further analysis, without loss of generality, we assume that $\tau=0.$

Let $\mu$ be a Borel measure defined on the Borel sets of the interval $(0,1).$
The distributed order differential operator with the mixing measure $\mu,$ by definition, is
\[
\mathcal{D}_{\mu,t}g(t)=\int_0^1 \mathcal{D}_{\ast}^{\beta} g(t) d \mu(\beta).
\]
Two  functions $\mathcal{K}_{\mu}(t)$ and $\Phi_{\mu}(s)$ defined with the help of the measure $\mu$ respectively by
\begin{align} \label{calk}
\mathcal{K}_{\mu}(t)&=\int_0^{1}\frac{t^{-\beta}}{\Gamma(1-\beta)}d\mu(\beta),
~ t>0,\\
\label{phi} \Phi_{\mu}(s)&=\int_0^1s^{\beta}d\mu(\beta), ~ s>0.
\end{align}
play an important role in our further analysis. We denote by  $ \tilde{g}(s)$ %
the Laplace transform of a function $g,$ that is
\[
\tilde{g}(s)=\int_0^{\infty}g(t)e^{-st}dt.
\]
  One can verify using the Laplace transform formula for fractional derivatives,
\begin{align}
\label{laplace} \widetilde{[\mathcal{D}_{\ast}^{\beta}g]}(s) &=
s^{\beta} \tilde{g}(s) - s^{\beta -1}g(0+)
\end{align}
that the formula %
\begin{equation}\label{Lapl1}
\widetilde{[\mathcal{D}_{\mu,t}g]}(s)=\Phi_{\mu}(s)
\tilde{g}(s)-g(0+)\frac{\Phi_{\mu}(s)}{s}, \, s>0,
\end{equation}
holds. Moreover, using the formula
 $\widetilde{[t^{-\beta}]}(s)=\Gamma(1-\beta)s^{\beta-1}$  (see \cite{Umarov_book}), one
can also verify that the relation
\begin{equation}
\widetilde{\mathcal{K}_{\mu}}(s)=\frac{\Phi_{\mu}(s)}{s}, \, s>0,
\end{equation}
holds.

The widest class of processes for which the It\^o calculus can be
extended is the class of semimartingales \cite{Protter}. A semimartingale is a
c\`adl\`ag process (the processes with paths which are right continuous and have left limits) which has the representation (see details in \cite{Protter})
$Z_t=Z_0+V_t+M_t,$
where $Z_0$ is a $d$-dimensional random vector,  $V_t$ is an adapted finite variation process, and $M_t$ is a
local martingale.
An example of  a semimartingale is a L\'evy process defined as follows. 
{A stochastic process $L_t \in \re^d, \ t\ge 0,$ is called a L\'evy process if the following three conditions are verified: (1) $L_0=0;$ (2) $L_t$ has independent stationary increments; (3) for all $\epsilon, t >0$, $\lim_{s\to t}
\mathbb{P}(|L_t-L_s|>\epsilon)=0.$ 
L\'evy processes have   c\`adl\`ag modifications. L\'evy processes
are fully described by three parameters: a vector $b \in
\mathbb{R}^n,$ a nonnegative definite matrix $\Sigma,$ and a measure
$\nu$ defined on $\mathbb{R}^n \setminus \{0\}$ such that $\int
\min (1,|x|^2) d \nu <\infty$, called its L\'evy measure. The
L\'evy-Khintchine formula provides a characterization of L\'evy process  through its characteristic
function $\Phi_t(\xi)=e^{t \Psi (\xi)}$, with the {L\'evy symbol}  (see, e.g. \cite{Sato})
\begin{equation}
\label{levy-khintchin} \Psi(\xi) = i(b,\xi) - (\Sigma
\xi, \xi) + \int_{\mathbb{R}^n \setminus \{0\}} (e^{i(x,\xi)}- 1-i (x,\xi)
1_{(|x| \le 1)})\nu(dx).
\end{equation}
A L\'evy subordinator is a nonnegative and nondecreasing L\'evy process. A L\'evy subordinator with the L\'evy symbol $\Psi(s)=s^{\beta}, \, s>0,$ is called a $\beta$-stable L\'evy subordinator.
The composition of two L\'evy processes is again a L\'evy process. In
particular, a time-changed
process $L_{D_t}$ with a stable L\'evy subordinator $D_t$ is again a
L\'evy process. Transition probabilities of such a time-changed
process satisfy FPK type equation with a pseudo-differential operator on
the right hand side, whose symbol is continuous but not smooth. However, the composition $L_{W_t},$ where
$W_t=\inf\{\tau \ge 0: D_{\tau} > t\},$ the inverse to the stable
subordinator $D_t$ is not a L\'evy process, but still is a
semimartingale. This drastically changes the associated FPK
equation: now it is a time-fractional pseudo-differential
equation \cite{HKU-1,HU}, implying non-Markovian behaviour of the
process.

The properties of the density function of the inverse process $W_t$ are listed below 
in the more general case when $W_t$
represents the inverse of a process which belongs to the class
$\mathcal{S}_{\mu}$ of mixtures of an arbitrary number of independent stable
subordinators with a mixing measure $\mu.$ 
Namely, a non-negative increasing L\'evy stochastic process $D_t$ belongs to the class $\mathcal{S}_{\mu}$ if 
\[
\ln \mathbb{E}\left[  e^{-sD_t} \right] = -t \Phi_{\mu}(s).
\]
It is known (see e.g. \cite{Sato}) that mixtures of independent
stable processes of different indices are no longer stable. For a stochastic process in $\mathcal{S}_{\mu}$ we use the notation $D_{\mu, t}$ showing the dependence on the measure $\mu,$ and for their inverses the notation $W_{\mu, t}.$
The density function of $W_{\mu, t}$ is denoted by $f^{\mu}_t(\tau).$ Note that such mixtures model complex diffusions and other types of stochastic processes with several simultaneous
diffusion modes. Their associated FPK type equations are
\textit{distributed order differential equations} (see e.g.
\cite{Kochubei,MSh,UG}).

\begin{lemma} \label{lemgen}The density function $f^{\mu}_t(\tau)$
possesses the following properties:
\begin{enumerate}
\item[$(a)$] $\lim_{t\to +0}f^{\mu}_t(\tau)=\delta_0(\tau), ~ \tau \ge 0;$\vspace{0.8mm}
\item[$(b)$] $\lim_{\tau \to +0}f^{\mu}_t(\tau)= \mathcal{K}_{\mu}(t),
~ t>0;$\vspace{0.8mm}
\item[$(c)$] $\lim_{\tau \to \infty}f^{\mu}_t(\tau) = 0, ~ t \ge 0;$\vspace{0.8mm}
\item[$(d)$] $\mathcal{L}_{t \to s}[f^{\mu}_t(\tau)](s)= \frac{\Phi_{\mu}(s)}{s} e^{-\tau \Phi_{\mu}(s)},
~ s>0, ~ \tau \ge 0.$
\end{enumerate}
\end{lemma}

\begin{lemma} \label{lemmaingen}
The function $f^{\mu}_t(\tau)$ % 
satisfies for each $t>0$ the equation
\begin{equation}
\label{relation3} \mathcal{D}_{\mu,t}f^{\mu}_{t}(\tau)=
-\frac{\partial}{\partial \tau} f^{\mu}_t(\tau) -\delta_0 (\tau)
\mathcal{K}_{\mu}(t),
\end{equation}
in the sense of distributions.
\end{lemma}

 For the proofs of these lemmas we refer the reader to \cite{Umarov_book}.

%%%%%%%%%%%%%%%%%%%%%%%%%%%%%%%%%%%%%%%%%%%%%%%%%%%
%%%%%%%%%%%%%%%%
%%%%%%%%%%%%%%%%%
\section{Fractional FPK equations associated with SDEs in a bounded domain}
\begin{theorem}
\label{thm_010}
Let $X_t$ be a stochastic process associated with the FPK equation \eqref{bound_500}-\eqref{bound_502} and  $W_{\mu, t}$ be the inverse to a L\'evy's subordinator $D_{\mu, t} \in \mathcal{S}_{\mu}$ with a mixing measure $\mu.$ Then the FPK equation associated with the time changed stochastic process $X_{W_{\mu, t}}$ has the form
\begin{align}
\mathcal{D}_{\mu, t} v(t,x)  = A(x,D) v(t,x), \quad & t>0, \ x \in \Omega,
\label{bound_600} 
\\
\mathcal{W}(x^{\prime}, D) v(t, x^{\prime})  =0, \quad & t>0, \ x^{\prime} \in \partial \Omega,
\label{bound_601}
\\
v(0,x) =u_0(x), \quad & x \in \Omega.
\label{bound_602}
\end{align} 
where $u_0 \in C^2_{\mathcal{W}}(\Omega) \equiv \{\vf \in C^2(\Omega): \mathcal{W}(x^{\prime}, D) \vf(x^{\prime})=0\}.$ 
\end{theorem}
{\it Proof.}
Let $T_t$ be the semigroup with the infinitesimal generator
$
A=A(x,D): C(\overline{\Omega}) \to C(\overline{\Omega}),
$
with the domain 
\[
\mathbb{D} (A) =\{\phi \in C^2(\Omega): \mathcal{W}(x^{\prime},D) \phi (x^{\prime})=0, \ x^{\prime} \in \partial \Omega\}.
\]
Then the unique solution of problem \eqref{bound_500}-\eqref{bound_502} has the form
\[
u(t,x)=T_t u_0(x)=\mathbb{E} [u_0(X_t)|X_0=x],
\]
indicating connection of the solution $u(t,x)$ of the FPK equation in \eqref{bound_500}-\eqref{bound_502} with the stochastic process $X_t.$ Now, consider the function $v(t,x)$ obtained from the latter replacing $X_t$ by $X_{W_{\mu, t}},$ that is
\begin{align}
v(t,x) & =\mathbb{E}[u_0(X_{W_{\mu, t}})|X_0=x]  
\notag \\ 
	 & = \int_0^{\infty} \mathbb{E} [u_0(X_{W_{\mu, t}})|W_{\mu, t}=\tau|X_0=0] d\tau  
\notag \\
         & = \int_0^{\infty} u(\tau,x) f_{W_{\mu, t}}(\tau)d\tau   
 \notag \\
         & = \int_0^{\infty} f_{W_{\mu, t}}(\tau) T_{\tau} u_0(x) d\tau.
         \label{fr_sln}
\end{align}
We will show that $v(t,x)$ defined above satisfies the initial-boundary value problem for fractional FPK equation in \eqref{bound_600}-\eqref{bound_602}.  First, we show that $v(t,x)$ satisfies equation \eqref{bound_600}.
Indeed, one can readily
see that 
\[
\displaystyle v(t, x)= -  \int_0^{\infty}
\frac{\partial}{\partial \tau} \{J
f_{_{D_{\mu, \tau}}}(t)\} \hspace{0.3mm}(T_{\tau} u_0(x)) d\tau.
\] 
Now it follows from the definition of $D_{\mu, t}$ that the Laplace transform of
$v(t,x)$ satisfies
\begin{align} \label{lapgen}
\tilde{v}(s,x) &= \frac{\int_0^1 s^{\beta} d\mu(\beta)}{s}  \int_0^{\infty}
e^{-\tau \int_0^1 s^{\beta}d\mu(\beta)}(T_{\tau}u_0(x))d\tau \notag 
\\
&=
\frac{\Phi_{\mu}(s)}{s} \hspace{0.5mm} \tilde{u}(\Phi_{\mu}(s),x), \quad s
> \bar{\omega},
\end{align}
where 
$\tilde{u}(s,x)$ is the Laplace transform of $u(t,x),$ the function $\Phi_{\mu}(s)$ is defined in \eqref{phi}, and
$\bar{\omega} >0$ is a number such that $s > \bar{\omega}$ if $\Phi_{\mu}(s) > \omega$
($\bar{\omega}$ is uniquely defined, since as is seen from Definition \eqref{phi}, the function $\Phi_{\mu}(s)$ is a strictly increasing function). The function $\tilde{u}(s,x)$ satisfies the equation 
\begin{equation} \label{eqn_010}
s \tilde{u}(s,x) - A(x,D) \tilde{u}(s,x)=u_0(x), \quad x\in \Omega,
\end{equation} 
Indeed, applying the Laplace transform to both sides of equation \eqref{bound_500} and taking into account the initial condition \eqref{bound_502}, we obtain equation \eqref{eqn_010}.
It follows from equations \eqref{lapgen} and \eqref{eqn_010} that the composit function
\[
\tilde{u}(\Phi_{\mu}(s),x)=s\tilde{v}(s,x)/\Phi_{\mu}(s)
\] 
satisfies the equation 
\[
\Big(\Phi_{\mu}(s) - A(x,D) \Big)\frac{s \tilde{v}(s,x)}{\Phi_{\mu}(s)} = u_0(x), \quad s > \bar{\omega}, \ x \in \Omega,
\]
or, the same, the equation
\begin{equation*}
\Big(\Phi_{\mu}(s)- A(x,D) \Big) \tilde{v}(s, x)= u_0 (x) \ \frac{\Phi_{\mu}(s)}{s}, \quad s > \bar{\omega}, x \in \Omega.
\end{equation*}
We rewrite the latter in the form
\begin{equation}
\label{new}
\Phi_{\mu}(s)\tilde{v}(s, x) - \frac{\Phi_{\mu}(s)}{s} v(0+,x) = A(x, D) \tilde{v}(s, x), \quad s > \bar{\omega}, \ x \in \Omega.
\end{equation}
We notice that the left hand side of the latter equation is the Laplace transform of the expression $\mathcal{D}_{\mu, t} v(t,x)$ due to formula \eqref{laplace}.  Therefore, equation \eqref{new} is equivalent to equation \eqref{bound_600}.

Further, using \eqref{bound_501}, we have
\begin{align*}
\mathcal{W}(x^{\prime},D) v (t, x^{\prime})&= \mathcal{W}(x^{\prime},D) \int_0^{\infty} u(\tau, x^{\prime}) f_{W_{\mu, t}}(\tau) d\tau
\\ &= \int_0^{\infty} \mathcal{W}(x^{\prime},D) u (\tau, x^{\prime}) f_{W_{\mu, t}}(\tau) d\tau
=
0
\end{align*}
since $W(x^{\prime},D) u (\tau, x^{\prime})=0$ for all $\tau >0$ due to boundary condition \eqref{bound_501}. 

Finally, making use of the dominated convergence theorem,
\begin{align*}
v(0,x) &= \lim_{t \to 0+} \mathbb{E}[u_0(X_{W_{\mu, t}})|X_{W_{\mu, 0}}=x] 
\\ &= \mathbb{E}[u_0(X_{W_{\mu, t}})|X_{W_{\mu, 0}}=x] 
\\& =\mathbb{E}[u_0(X_0)|X_{0}=x] 
\\& =\mathbb{E}[u_0(x)] = u_0(x).
\end{align*}
Hence, $v(t,x)$ defined in \eqref{fr_sln} satisfies the initial-boundary value problem in \eqref{bound_600}-\eqref{bound_602} for the fractional order FPK equation.
\eproof

\medskip

In the particular case of $W_t$ being the inverse of a single L\'evy's subordinator $D_t$ with the stability index $\beta \in (0,1)$ this theorem implies the following result:
\begin{cor}
\label{cor_010}
Let $X_t$ be a stochastic process associated with the FPK equation \eqref{bound_500}-\eqref{bound_502} and  $W_t$ be the inverse to the L\'evy's stable subordinator with a stability index $0<\beta<1.$ Then the FPK equation associated with the time changed stochastic process $X_{W_t}$ has the form
\begin{align}
&\mathcal{D}_{\ast}^{\beta} v(t,x)   =  A(x,D) v(t,x), \quad  t>0, \ x \in \Omega,
\label{bound_700} 
\\
&\mathcal{W}(x^{\prime}, D) v(t, x^{\prime})   = 0, \quad  t>0, \ x^{\prime} \in \partial \Omega,
\label{bound_701}
\\
&v(0,x)   = u_0(x), \quad  x \in \Omega.
\label{bound_702}
\end{align} 
\end{cor}

An important question is the existence of a unique solution of the initial-boundary value problem in equations \eqref{bound_700}-\eqref{bound_702}. 

\begin{theorem}
\label{uniqueness_1}
Let the conditions $(i)-(iv),$ $(a)-(d),$ and $(C1)-(C3)$ are verified. Then initial-boundary value problem \eqref{bound_700}-\eqref{bound_702} for fractional order FPK equation has a unique solution $v(t,x)$ in the space  $C((0,\infty) \times \bar{\Omega})\cap C^{1}(t>0; C^{2}_{\mathcal{W}}(\Omega).$
\end{theorem}
\begin{remark}
Here $C^{1}(t>0; C^{2}_{\mathcal{W}}(\Omega)$ is the space of vector-functions differentiable in $t$ and belonging to $C^{2}_{\mathcal{W}}(\Omega)$ for each fixed $t>0.$
\end{remark}
{\it Proof.}
Introduce the operator $\mathfrak{U}_{\mathcal{W}}$ as follows: $\mathfrak{U}_{\mathcal{W}}= A(x,D)$ with the domain $Dom(\mathfrak{U}_{\mathcal{W}})=\{\phi \in C(\bar{\Omega}): A(x, D) \phi \in C(\bar{\Omega}), \mathcal{W}(x^{\prime}, D) \phi (x^{\prime})=0, \forall x^{\prime} \in \partial \Omega\}.$ As is proved in \cite{Taira-1} if the conditions of the theorem are verified, then 
there exists  a Feller semigroup $\{T_t\}_{t \ge 0}$ (nonnegative and contractive) on $\bar{\Omega}$ generated by $\mathfrak{U}_{\mathcal{W}}.$ That is for any $\vf \in C(\bar{\Omega})$ such that $0 \le \vf(x) \le 1$ on $\bar{\Omega}$ one has $0 \le T_t\vf(x) \le 1$ on $\bar{\Omega}.$ Moreover, it follows from the general semigroup theory (see, e.g. \cite{Taira}) that the equation $T_t \vf(x) = e^{t\mathfrak{U}_{\mathcal{W}}} \vf(x)$ holds. Hence, for an arbitrary $u_0 \in C(\bar{\Omega})$ the function $u(t,x) = T_t u_0 (x), \ t \ge 0, x \in \Omega,$ exists and solves the following initial value problem for a differential-operator equation
\begin{align}
\frac{\partial u(t,x)}{\partial t} &= \mathfrak{U}_{\mathcal{W}} \ u(t, x), \quad t \ge 0, \ x \in \Omega,
\\
\lim_{t \to 0} u(t, x) &= u_0(x), \quad x \in \Omega.
\end{align}
Since the operator $A(x, D)$ is elliptic with the spectrum in the negative real axis, it follows from the smoothness of a solution to parabolic equations  that $u(t,x)$ has all the derivatives if $t>0$ and $x \in \Omega.$ Thus, in particular,
this function belongs to the space  $C((0,\infty) \times \bar{\Omega})\cap C^{1}(t>0; C^{2}_{\mathcal{W}}(\Omega).$ 
%if $\vf \in C(\bar{\Omega}).$ 
The existence and uniqueness of a solution of initial-boundary value problem \eqref{bound_600}-\eqref{bound_602} immediately follows from representation  \eqref{fr_sln}. Also, it follows from this representation  that $v(t,x)$ has all derivatives if $t>0$ and that the estimate
\begin{align*}
|v(t,x)| & \le \int_0^{\infty} f_t^{\mu} (\tau) |T_{\tau} u_0(x)| d\tau \le  \sup_{t \ge 0} \|T_{t} u_0 (x)\|_{C(\bar{\Omega})} \\
& = \sup_{t \ge 0} \|u(t,x) \|_{C(\bar{\Omega})}, \quad t \ge 0, \ x \in \bar{\Omega},
\end{align*}
holds. 
Thus, the function $v(t,x)$ inherits all the properties of $u(t,x),$ including  the property $C((0,\infty) \times \bar{\Omega})\cap C^{1}(t>0; C^{2}_{\mathcal{W}}(\Omega).$
\eproof

%\begin{remark} (On non transversal case of boundary conditions) Suppose that the following two conditions are verified:
%\begin{enumerate}
%\item  $\mu(x^{\prime})+\delta(x^{\prime}) - \gamma(x^{\prime}) >0$ for all $x^{\prime} \in \partial \Omega,$
%\item $\mu(x^{\prime}) $ Then
%\end{enumerate}
%\end{remark}
\medskip

Now consider the following initial-boundary value problem \eqref{bound_500}-\eqref{bound_502} for $t$-dependent generalization of the FPK equation

\begin{align}
\frac{\partial u(t,x)}{\partial t}  = & \ B(x,D) u(t,x) + \frac{(\gamma+1)t^{\gamma}}{2}A(x,D) u(t,x), \quad  t>0, \ x \in \Omega,
\label{bound_700} 
\\
&\mathcal{W}(x^{\prime}, D) u(t, x^{\prime})  =0, \quad  t>0, \ x^{\prime} \in \partial \Omega,
\label{bound_701}
\\
&u(0,x) =\vf(x), \quad  x \in \Omega,
\label{bound_702}
\end{align} 
where $B(x,D)$ is a pseudo-differential operator whose order is strictly less than the order of the operator $A(x,D).$ We assume that $A(x, D)$ is an elliptic Waldenfels operator defined in \eqref{bound_5}
and $\mathcal{W}(x^{\prime}, D)$ is a Wentcel's boundary pseudo-differential operator defined in \eqref{bound_503}. The parameter $\gamma$ runs in the interval $(-1,1).$ Equation \eqref{bound_700} is a parabolic equation: if $0<\gamma<1,$ then it is degenerate; if $-1<\gamma<0,$ then it is singular. 
Obviously, the initial-boundary value problem \eqref{bound_700}-\eqref{bound_702} recovers  problem \eqref{bound_500}-\eqref{bound_502}, if $B(x, D) = 0$ and $\gamma=0.$

Here is the motivating example:  if $B(x, D)=0,$ $\gamma=2H-1$ and $A(x, D)=\Delta,$ Laplace operator, 
then equation (\ref{bound_700}) reduces to
\[
\frac{\partial u(t,x)}{\partial t} = Ht^{2H-1} \Delta u(t,x), \quad t>0, \ x\in \Omega,
\]
which represents the FPK equation associated with the fractional Brownian motion with the Hurst parameter $H \in (0,1)$ (see, e.g. \cite{Biagini,HKU-2}). Therefore, initial-boundary value problem \eqref{bound_700}-\eqref{bound_702} can be considered as a FPK equation of a stochastic process driven by fractional Brownian motion. By definition, the fractional Brownian motion with the Hurst parameter $H$ is a  stochastic process $B_t^{H},$ Gaussian with mean zero for each fixed $t \ge 0,$ with continuous paths and the following covariance function:
\[
R_H(s,t)=\mathbb{E}(B_t^H B_s^H) = \frac{1}{2} \left( t^{2H} +s^{2H} -|t-s|^{2H}  \right).
\]
Fractional Brownian motion is not Markovian and is not a semimartingale. Therefore, a stochastic integral and stochastic differential equation driven by a fractional Brownian motion can not be defined in the It\^o sense. Nevertheless, one can define stochastic differential equations driven by fractional Brownian motion in a meaningful sense \cite{Bender,DU,Nualart}. Thus, the class of initial-boundary value problems of the form \eqref{bound_700}-\eqref{bound_702} describes stochastic processes in a bounded region $\Omega$ driven by not only L\'evy processes, but also fractional Brownian motion.
Below we derive the fractional FPK equation associated with such a stochastic process with a time-changed driving process.

\begin{theorem}
\label{thm_050}
Let $X_t$ be a stochastic process associated with the FPK equation \eqref{bound_700}-\eqref{bound_702} and  $W_{\mu, t}$ be the inverse to a L\'evy's subordinator $D_{\mu, t} \in \mathcal{S}_{\mu}$ with a mixing measure $\mu.$ Then the FPK equation associated with the time changed stochastic process $X_{W_{\mu, t}}$ has the form
\begin{align}
\mathcal{D}_{\mu, t} v(t,x)  = &B(x, D)v(t,x) + \frac{\gamma + 1}{2} G_{\mu, \gamma, t} A(x,D) v(t,x), \quad  t>0, \ x \in \Omega,
\label{bound_800} 
\\
&\mathcal{W}(x^{\prime}, D) v(t, x^{\prime})  =0, \quad  t>0, \ x^{\prime} \in \partial \Omega,
\label{bound_801}
\\
&v(0,x) =u_0(x), \quad  x \in \Omega.
\label{bound_802}
\end{align}  
where the operator $G_{\mu, \gamma,t}$ %acts on {the} variable $t$ and
{is} defined as
\begin{equation} \label{presentation1}
G_{\mu, \gamma, t} v(t,x)= \Phi_{\mu} (t)
 \ast
\mathcal{L}^{-1}_{s \to t}\big[\frac{\Gamma(\gamma+1)}{2\pi i}
\int_{C-i \infty}^{C+i\infty} \frac{ m_{\mu}(z)
\tilde{v}(z,x)}{(\rho(s)-\rho(z))^{\gamma+1}}dz \big](t),
\end{equation}
where $\ast$ denotes the convolution operation, the symbol $\mathcal{L}^{-1}_{s \to t}$ means the inverse Laplace transform, 
$0<C<s,$ the functions $\rho(z)$ and $m_{\mu}(z)$ are defined by
\[
\rho(z)= \int_0^1 e^{\beta \mbox{Ln} (z)}d\mu(\beta),
\quad
m_{\mu}(z)=\frac{\int_0^1 \beta z^{\beta} d \mu(\beta)}{\rho(z)},
\]
the function $\Phi_{\mu}(t)$ is defined in (\ref{phi}), 
and
$u_0 \in C^2_{\mathcal{W}}(\Omega) \equiv \{\phi \in C^2(\Omega): \mathcal{W}(x^{\prime}, D) \phi (x^{\prime})=0\}.$ 
\end{theorem}

\textit{Proof.} Let $u(t,x)$  and  $v(t,x)$ be density functions of $X_t$ and $X_{W_{\mu, t}},$ respectively.
Then due to the formula for the conditional density, we have 
\begin{align}
\label{def_505}
v(t,x) & =   f_{X_{W_{\mu, t}}} (x) \notag \\ 
&= \int_0^{\infty} f_{(X_{W_{\mu, t}|W_{\mu, t}=\tau})} \mathbb{P} (W_{\mu, t} \in d\tau) \notag \\ 
&=
\int_{0}^{\infty} u(\tau, x) \mathbb{P} (W_{\mu, t} \in d \tau)   \notag \\
&=\int_0^{\infty} f_t^{\mu} (\tau) u(\tau, x) d\tau, 
\quad t \ge 0, \ x\in \Omega.
\end{align}
By definition of $X_t$ the function $u(t,x)$ solves problem \eqref{bound_700}-\eqref{bound_702}. We will show that $v(t,x)$ satisfies problem \eqref{bound_800}-\eqref{bound_802}.
To show that $v(t,x)$ satisfies equation \eqref{bound_800}, we compute
\[
D_{\mu,t}v(t,x)=\int_0^{\infty} D_{\mu,t} {f^\mu_t}{(\tau)}u(\tau, x)d\tau.
\] 
Here the change of the
order of $D_{\mu,t}$ and the integral is valid thanks to the estimate obtained in \cite{HU} for the density function $f_t^{\mu}(\tau)$
%\eqref{est} extends to a 
of a mixture of stable subordinators having mixing measure $\mu$ with
$supp \, \mu \subset (0,1).$ It follows from Lemma \ref{lemmaingen} that
\begin{align} \label{thm11}
D_{\mu,t}v(t,x)=-\int_0^{\infty}
\frac{\partial{f^\mu_t(\tau)}}{\partial \tau}u(\tau)d\tau -
\mathcal{K}_{\mu}(t)\int_0^{\infty} \delta_0(\tau)u(\tau)d\tau.
\end{align}
Integration by parts in the first integral, we have %in \eqref{thm11} gives
\begin{equation}
\label{add_202}
-\int_0^{\infty}
\frac{\partial{f^\mu_t(\tau)}}{\partial \tau}u(\tau)d\tau
=
\int_0^{\infty} {f^\mu_t(\tau)}\frac{\partial u(\tau)}{\partial
\tau}d\tau - \lim_{\tau \to
\infty}{f^\mu_t(\tau)}u(\tau) + \lim_{\tau \to
0}{f^\mu_t(\tau)}u(\tau).
\end{equation}
The first limit on the right hand side is zero due to part (c)
of Lemma \ref{lemgen}. %and the condition on the spectrum of the operator $A.$ 
Due to part (b) of Lemma \ref{lemgen} the second limit on the right hand side of \eqref{add_202} has the same value as the second integral on the right side of \eqref{thm11}, but with the opposite sign. Hence, it follows that
\begin{align*}
D_{\mu,t}v(t,x)&= \int_0^{\infty} f^{\mu}_t(\tau)
\frac{\partial}{\partial
  \tau} u(\tau,x) d \tau.
\end{align*}
Now using equation \eqref{bound_700}, we have
\begin{align*}
D_{\mu,t}v(t, x) & = \int_0^{\infty} f_t^{\mu}(\tau) \left[ B(x,D) u(\tau,x) + \frac{(\gamma+1) \tau^{\gamma}}{2}A(x,D) u(\tau,x)  \right] d\tau            \\
&= B(x,D) \int_0^{\infty} f_t^{\mu}(\tau) u(\tau, x) d\tau + \frac{\gamma+1}{2} A(x, D) \int_0^{\infty} f_t^{\mu}(\tau) \tau^{\gamma} u(\tau, x) d \tau \\
&= B v(t, x) + \frac{\gamma + 1}{2} A (x, D) \int_0^{\infty} f^{\mu}_t(\tau)\tau^{\gamma} u(\tau, x)d\tau \\
&= B v(t, x) + \frac{\gamma + 1}{2} A G_{\mu, \gamma, t} v(t, x),
\end{align*}
where
\begin{equation} \label{gt}
G_{\mu, \gamma,t} v(t, x) = \int_0^{\infty}
f^{\mu}_t(\tau)\tau^{\gamma}u(\tau, x)d\tau.
\end{equation}
The fact that the operator $G_{\mu, \gamma, t}$ has the representation \eqref{presentation1} is proved in \cite{HKU-2}. 

Further, we show that  $\mathcal{W}(x^{\prime}, D) v(t, x^{\prime}) = 0 $ if $x^{\prime} \in \partial \Omega.$ 
%can be shown exactly as in the previous theorem. 
Indeed, using $\mathcal{W}(x^{\prime}, D) u(\tau, x^{\prime}) = 0$ for all $\tau \ge 0,$ we have
\begin{align*}
\mathcal{W}(x^{\prime},D) v (t, x^{\prime})  % \mathcal{W}(x^{\prime},D) \int_0^{\infty} u(\tau, x^{\prime}) f_{W_{\mu, t}}(\tau) d\tau \\
 &= \int_0^{\infty} \mathcal{W}(x^{\prime},D) u (\tau, x^{\prime}) f_{W_{\mu, t}}(\tau) d\tau
=
0, \quad x^{\prime} \in \partial \Omega,
\end{align*}
for any fixed $t \ge 0.$

Finally, making use of part (a)
of Lemma \ref{lemgen} and the dominated convergence theorem,
\begin{align*}
\lim_{t \to 0+} v(t,x) &= \lim_{t \to 0+} \int_0^{\infty} f_t^{\mu}(\tau) u(\tau, x) d \tau
\\ &= \int_0^{\infty} \lim_{t \to 0+}  f_t^{\mu}(\tau)  \ u(\tau, x) d \tau 
\\& =\int_0^{\infty} \delta_0(\tau)  \ u(\tau, x) d \tau  
\\& =u(0, x) = u_0(x).
\end{align*}
Hence, $v(t,x)$ defined in \eqref{def_505} satisfies the initial-boundary value problem in \eqref{bound_800}-\eqref{bound_802} for time dependent fractional order FPK equation.
\eproof

\begin{remark}
{\em
The properties of the
operator $G_{\gamma,t}$ are studied in paper \cite{HKU-2}, including
the fact that the family $\{G_{\gamma,t}, \, -1 <\gamma < 1\}$
possesses the semigroup property. Namely, for
any $\gamma, \delta \in (-1,1), \gamma+\delta \in (-1,1),$ one has
$G^{\mu}_{\gamma,t} \circ
G^{\mu}_{\delta,t}=G^{\mu}_{\gamma+\delta,t}=G^{\mu}_{\delta,t}
\circ G^{\mu}_{\gamma,t},$ where ``$\circ$" denotes the composition
of two operators.
}
\end{remark}

Similar to Corollary \ref{cor_010}, in the particular case of $W_t$ being the inverse of a single L\'evy's subordinator $D_t$ with the stability index $\beta \in (0,1)$ Theorem \ref{thm_050} implies the following result:
\begin{cor}
\label{cor_050}
Let $X_t$ be a stochastic process associated with the FPK equation \eqref{bound_700}-\eqref{bound_702} and  $W_t$ be the inverse to the L\'evy's stable subordinator with a stability index $0<\beta<1.$ Then the FPK equation associated with the time changed stochastic process $X_{W_t}$ has the form
\begin{align*}
&\mathcal{D}_{\ast}^{\beta} v(t,x)   =  B(x, D)v(t,x) + \frac{\gamma + 1}{2} G_{\mu, \gamma, t} A(x,D) v(t,x), \quad  t>0, \ x \in \Omega,
%\label{bound_700} 
\\
&\mathcal{W}(x^{\prime}, D) v(t, x^{\prime})   = 0, \quad  t>0, \ x^{\prime} \in \partial \Omega,
%\label{bound_701}
\\
&v(0,x)   =u_0(x), \quad  x \in \Omega.
%\label{bound_702}
\end{align*} 
\end{cor}

%Concluding, we briefly discuss the existence of the stochastic process associated with the fractional order FPK equations derived in Theorems \ref{thm_010} and \ref{thm_050} and Corollaries \ref{cor_010} and \ref{cor_050}. Our arguments are based on the representations \eqref{fr_sln} and \eqref{def_505}. 

\begin{theorem}
\label{uniqueness_fr}
Let the operators $A(x,D),$ $B(x,D),$ and $\mathcal{W}(x^{\prime}, D)$ in problem \eqref{bound_800}-\eqref{bound_802} satisfy the following conditions: 
\begin{enumerate}
\item[(A)] The pseudo-differential operator $A(x,D)$ is an elliptic Waldenfels operator defined in \eqref{bound_5} with the conditions $(i)-(iv)$ and $(C1)$ satisfied;
\item[(B)] The  operator $B(x,D)$ is a pseudo-differential operator whose order is strictly less then the order of $A(x, D);$
\item[(W)] The pseudo-differential operator $\mathcal{W}(x^{\prime}, D)$ is a Wentcel's boundary operator defined in \eqref{bound_503} with the conditions $(a)-(d),$  $(C2),$ and $(C3)$ satisfied.
\end{enumerate}
Then for an arbitrary $u_0 \in C^{2}_{\mathcal{W}} (\Omega)$ initial-boundary value problem \eqref{bound_800}-\eqref{bound_802} %for fractional order FPK equation 
has a unique solution $v(t,x)$ in the space  $C([0,\infty) \times \bar{\Omega})\cap C^{1}(t>0; C^2_{\mathcal{W}}(\Omega)).$ 
\end{theorem}

{\it Proof.}
The arguments used in the proof of Theorem \ref{uniqueness_1} do not work in the case of problem \eqref{bound_800}-\eqref{bound_802}, since the solution $u(t,x)$ of problem \eqref{bound_700}-\eqref{bound_702} does not have a representation through the Feller semigroup. 
%However, in this case one can use the existence theorem for general boundary value problems for parabolic equations; see \cite{Friedman,,}. Assume for each fixed $t \ge 0$ the function $u(t,x) \in C_{\mathcal{W}}(\Omega)$ and consider the following integral equation
%\begin{equation}
%u(t,x) = \vf(x) + B(x, D) \int_0^t u(s,x)ds + \frac{\gamma + 1}{2}A(x,D) \int_0^t s^{\gamma} u(s,x) ds, \quad t \ge 0, \ x \in \Omega, 
%\end{equation}
To prove the theorem we will use the properties of pseudo-differential operators.

Consider the operator 
\[
\mathcal{A}_{\mathcal{W}}(t) = \frac{t^{\gamma}}{2} A(x,D) +  B(x,D)
\] 
with the domain $Dom(\mathcal{A}_{\mathcal{W}}(t))=\{\phi \in C^{2}(\overline{\Omega}): \mathcal{W}(x^{\prime},D) \phi(x^{\prime})=0, \ x^{\prime} \in \partial \Omega\}.$ This operator is a pseudo-differential operator with the symbol 
\begin{equation}
\label{sigma0}
\sigma(t, x, \xi) = \frac{t^{\gamma }}{2} \sigma_A(x, \xi) +  \sigma_B(x, \xi), \quad t\ge 0, x \in \Omega, \ \xi \in \re^d,
\end{equation}
where
$\sigma_A(x, \xi)$ and $\sigma_B(x, \xi)$ are symbols of the operators $A(x, D)$ and $B(x, D),$ respectively.
Due to conditions $(A)$ and $(B)$ of the theorem, for each fixed $t > 0$
the symbol $\sigma(t, x,\xi)$ satisfies the following ellipticity estimate
\begin{equation}
\label{sigma}
-\sigma(t, x, \xi) \ge  \kappa_{t} |\xi|^{\delta}, \quad |\xi| \ge C,
\end{equation}
with some constants $\kappa_t>0,$ $C>0,$ and $\delta>0.$

One can see that the solution of problem \eqref{bound_700}-\eqref{bound_702} has a formal representation
\begin{equation}
\label{solution_000}
u(t,x) = \mathcal{S}(t,x,D) u_0(x), \quad t \ge 0, \ x \in \Omega,
\end{equation}
where the solution pseudo-differential operator $\mathcal{S}(t,x,D)$ has the symbol
\begin{equation}
\label{solution_001}
{s}(t,x,\xi)=e^{ t  \sigma(t,x,\xi)}, \quad t \ge 0, \ \xi \in \re^d.
\end{equation}
The fact that $u(t,x)$ satisfies equation \eqref{bound_700} can be verified by direct calculation. To show this fact let us extend $u_0(x)$ for all $x \in \re^d \setminus \Omega$ by zero, and denote the extended function again by $u_0(x).$ Let $\tilde{u}_0(\xi)$ is the Fourier transform of $u_0(x).$ Then, we have
\begin{align*}
\frac{\partial u(t,x)}{\partial t} &= \frac{\partial \mathcal{S}(t,x,D) u_0(x)}{\partial t} \\
&= \frac{1}{(2\pi)^d} \frac{\partial}{\partial t} \int\limits_{\re^d}  e^{ t  \sigma(t,x,\xi)} e^{-ix \xi} \tilde{u}_0(\xi) d \xi \\
&= \frac{1}{(2\pi)^d}  \int\limits_{\re^d} \left[ \sigma(t,x,\xi) + t \ \frac{\partial \sigma(t,x,\xi)}{\partial t}  \right] e^{ t  \sigma(t,x,\xi)} e^{-ix \xi} \tilde{u}_0(\xi) d \xi 
\\
&= 
\frac{1}{(2\pi)^d}  \int\limits_{\re^d} \left[ \frac{(\gamma + 1)t^{\gamma}}{2} \sigma_{A}(x,\xi) + \sigma_B(x,\xi) \right] e^{-ix \xi} e^{ t  \sigma(t,x,\xi)} \tilde{u}_0(\xi) d \xi 
\\
&= 
\left[ \frac{(\gamma + 1)t^{\gamma}}{2} A(x, D) + B(x, D) \right] w(t,x), \quad t>0, \ x \in \Omega,
\end{align*}
where $w(t,x)$ has the Fourier transform
\[
\tilde{w}(t, \xi) = e^{ t  \sigma(t,x,\xi)} \tilde{u}_0(\xi), \quad t>0, \ \xi \in \re^d.
\]
Changing the differentiation and integration operators in the above calculation is legitimate due to estimate \eqref{sigma}. 
Now calculating the inverse Fourier transform of $\tilde{w}(t,\xi)$, and using the definition \eqref{psdo0} of pseudo-differential operators, we have
\begin{align*}
w(t,x) &= \frac{1}{(2\pi)^d}  \int\limits_{\re^d}  e^{ t  \sigma(t,x,\xi)} e^{-ix \xi} \tilde{u}_0(\xi) d \xi 
\\
&=  \frac{1}{(2\pi)^d}  \int\limits_{\re^d}  s(t, x, \xi)e^{-ix \xi} \tilde{u}_0(\xi) d \xi \\
&= \mathcal{S}(t,x,D) u_0(x), \quad t>0, \ x \in \Omega.
\end{align*}
Thus, $w(t,x) = u(t,x),$ and hence $u(t,x)$ defined by \eqref{solution_000} satisfies equation \eqref{bound_700}.
Further, since $\mathcal{W}(x^{\prime},D)u_0(x^{\prime})=0$ for $x^{\prime} \in \partial \Omega,$ it follows from \eqref{solution_000} that $\mathcal{W}(x^{\prime},D)u(t,x^{\prime})=0$ for all $t >0$ and $x^{\prime} \in \partial \Omega.$ Moreover, it follows from \eqref{solution_001} that $\mathcal{S}(0,x, D) = I,$ the identity operator, implying $u(0,x)=u_0(x).$
Finally, estimate \eqref{sigma} and representation \eqref{solution_000} of the solution imply 
the inclusion of the solution to the space  $C([0,\infty) \times \bar{\Omega})\cap C^{1}(t>0; C^2_{\mathcal{W}}(\Omega)).$ 

The existence and uniqueness of a solution to initial-boundary value problem \eqref{bound_800}-\eqref{bound_802} immediately follows from representation  \eqref{def_505}, namely
\[
v(t,x)=\int_{0}^{\infty} f_t(\tau) u(\tau, x) d \tau,
\]
where $u(t,x)$ is the solution of problem \eqref{bound_700}-\eqref{bound_702}. The arguments here are similar to the proof of Theorem \ref{uniqueness_1}. Also, similar to Theorem \ref{uniqueness_1}, the function $v(t,x)$ inherits all the properties of $u(t,x),$ including  the property $C((0,\infty) \times \bar{\Omega})\cap C^{1}(t>0; C^{2}_{\mathcal{W}}(\Omega).$
\eproof

\begin{remark}
The technique used to prove Theorem \ref{uniqueness_fr} is applicable for FPK equations in the whole space $\re^d$ obtained in papers \cite{HU,HKU-2}.
\end{remark}

%\vspace{1cm}
%{\bf Semigroup property, existence of a solution, fractional Feinman-Kac, etc. }

%\section*{References}


\begin{thebibliography}{99}


\bibitem{Agrawal}
	Agrawal, O.P. (2002). Solution for a fractional diffusion wave equation in a bounded domain.
	\textit{Nonlinear Dynamics,} {\bf 29}:145--155.


\bibitem{BC}
	Baudoin, F., Coutin, L. (2007). Operators associated with a stochastic differential equations driven by fractional Brownian motion. \textit{Stoch. Process. Appl.}, {\bf 17} (5):550--574. 


\bibitem{BN}
	Barndorff-Nielsen, O.E., Mikosch, T, Resnick, S. (eds). (2001).
	\textit{L\'evy Processes: Theory and Applications}, Birkh\"auser.

\bibitem{Bender}
	Bender, C. (2003). An It\^o formula for generalized functionals of a fractional Brownian motion with arbitrary Hurst parameter.
	\textit{Stochastic Processes and Their Applications}, {\bf 104} (1):81--106.


\bibitem{Biagini}
	Biagini, F., Hu, Y., Oksendal, B., Zhang, T. (2008). 
	\textit{Stochastic Calculus for Fractional Brownian Motion and Application.} Springer.


\bibitem{Bogolyubov}
	Bogolyubov, N.N., Sankovich ,D.P. (1994).
	{N.N. Bogolyubov and statistical mechanics.} 
	 \textit{Russian Math. Surveys,} {\bf 49} (5):19--49.

\bibitem{Bressloff}
              Bressloff, P.C. (2014).
              \textit{Stochastic Processes in Cell Biology}. Springer.
              
\bibitem{Burch}
	Burch, N., Lehouch, R. (2011).
	Classical, nonlocal, and fractional diffusion equations on bounded domains.
	\textit{International Journal for Multiscale Computational Engineering,} {\bf 9} (6): 661--674. 


\bibitem{Chechkin}
	Chechkin, A.V., Klafter, J., Sokolov, I.M. (2003).
	Fractional Fokker-Planck equation for ultraslow kinetics.
	\textit{Europhysics Letters,} \textbf{63} (3):326--332.
	
\bibitem{MMM-1}
	Chen, Zh.-Q., Meerschaert, M., Nane, E. (2012). 
	Space-time fractional diffusion on bounded domain.
	\textit{J. Math. Anal. Appl.} {\bf 393}:479--488.

\bibitem{DU}
	Decreusefond, L., \"Ust\"unel, A.S. (1998). 
	Stochastic analysis of the fractional Brownian motion. 
	\textit{Potential Analysis}, {\bf 10} (2):177--214.


\bibitem{Gajda}
	Gajda, J. (2013).
	Fractional Fokker-Planck equation with space dependent drift and diffusion: the case of tempered $\alpha$-stable waiting times.
	\textit{Acta Physica Polonica B.} {\bf 44}:1149--1161.


\bibitem{Gardiner}
	Gardiner, C.W. (1985).
	\textit{Handbook of Stochastic Methods: for Physics, Chemistry and the Natural Sciences.} Springer.

\bibitem{HKU-1}
	Hahn, M., Kobayashi, K., Umarov, S. (2012).
	SDEs driven by a time-changed L\'evy process and their time-fractional order pseudo-differential equations. 
	\textit{J. Theoret. Probab.} {\bf 25} (1):262--279.
	
	
\bibitem{HU} 
	Hahn, M., Umarov, S. (2011). 
	Fractional Fokker-Planck-Kolmogorov type equations and their associated stochastic differential equations. 
	 \textit{Frac. Calc. Appl. Anal.} {\bf 14} (1):56--79.

\bibitem{HKU-2} 
	Hahn, M, Kobayashi, K., Umarov, S. (2011).
	 Fokker-Planck-Kolmogorov equations associated with time-changed fractional Brownian motion.
	\textit{Proc. Amer. Math. Soc.} {\bf 139} (2):691--705.
	

\bibitem{Hormander}
	H\"ormander, L. (1965).
	Pseudo-differential operators. \textit{Comm. Pure Appl. Math.} {\bf 18}:501--517.

\bibitem{Jacob}
	Jacob, N. (2001).
	\textit{Pseudo-differential Operators and Markov Processes}, Imperial College Press, London.


\bibitem{Kochubei}
	Kochubei, A. (2008).
	Distributed order calculus and equations of ultraslow diffusion.
	\textit{J. Math. Anal. Appl.} {\bf 340} (1):252--281.

\bibitem{KohnNirenberg}
	Kohn, J. J., Nirenberg, L. (1965).
	An algebra of pseudo-differential operators. \textit{Comm. Pure Appl. Math. } {\bf 18}:269--305.

\bibitem{MMM-2} 
	Meerschaert, M., Nane, E., Vellaysamy, P. (2009). 
	Fractional Cauchy problems on bounded domain.
	\textit{Annals of Probability,} \textbf{37} (3):979--1007.
	
\bibitem{MSh}
	Meerschaert, M., Scheffler, H.-P. (2006).
	Stochastic model for ultraslow diffusion.
	\textit{Stochastic Processes and Their Applications}, {\bf 116} (9):1215--1235.
	

\bibitem{Melnikov}
	Melnikov, A.V. (1999).
	Financial Markets: Stochastic Analysis and the Pricing of Derivative Securities.
	\textit{Translations of Mathematical Monographs}, {\bf 184}, AMS. 


\bibitem{MetzlerKlafter00}
     Metzler, R., Klafter, J. (2000).
     The random walk's guide to anomalous diffusion: a fractional dynamics approach.
     \textit{Phys. Rep}. \textbf{339} (1):1--77.

\bibitem{Nualart}
	Nualart, D. (2006). 
	\textit{The Malliavin Calculus and Related Topics}, (2nd ed.), Springer.

\bibitem{Presse}
	Presse, S. (2015).
	A data-driven alternative to the fractional Fokker-Planck equation.
	\textit{Journal of Statistical Mechanics: Theory and Experiment.} {\bf 2015}, July.

\bibitem{Protter}
	Protter, P. (1991).
	\textit{Stochastic Integration and Differential Equations.} Springer-Verlag.

\bibitem{Sato}
    {Sato, K-i.} (1999).
    \textit{L\'evy Processes and Infinitely Divisible Distributions.}
    Cambridge University Press.

\bibitem{Shizgal}
           Shizgal, B. (2015). 
           \textit{Spectral Methods in Chemistry and Physics: Applications to Kinetic Theory and Quantum Mechanics.} Springer.

\bibitem{Taira}
	Taira, K. (2014). 
	\textit{Semigroups, Boundary Value Problems and Markov Processes.}
	Springer.

\bibitem{Taira-1}
	Taira, K. (1992). 
	On the existence of Feller semigroups with boundary conditions.
	\textit{Memoirs of the AMS}, {\bf 475} (2).

\bibitem{Tomovski}
	Tomovski, Z., Sandev, T. (2013).
	Exact solutions for fractional diffusion equation in a bounded domain with different boundary conditions.
	\textit{Nonlinear Dynamics,} {\bf 71} (4): 671--683.


\bibitem{Umarov_book} 
	Umarov, S. (2015).
	\textit{Introduction to Fractional and Pseudo-Differential Equations with Singular Symbols.}
	Springer.


\bibitem{UG}
	Umarov, S., Gorenflo, R. (2005).
	The Cauchy and multi-point problem for distributed order fractional differential equations.
	\textit{ZAA}, {\bf 24}:449--466.

\bibitem{Kampen} 
	Van Kampen, N.G. (1997).
	\textit{Stochastic Processes in Physics and Chemistry.} 
	Elsevier.	

\bibitem{Wentcel}
	Wentcel, A.D. (1959).
	On boundary conditions for multidimensional diffusion processes. {\it Theory of Probability and its Applications}, {bf 4}:164--177. 

\bibitem{Weron} 
	Weron, K., Magdziarz, M. (2008). 
	Modeling of subdiffusion in space-time-dependent force fields beyond the fractional Fokker-Planck equation.
	\textit{Physical Review E,} {\bf 77}, 036704.
	




\end{thebibliography}
\end{document}